\documentclass[fontsize=12pt]{article}
\usepackage[utf8]{inputenc}
\usepackage{color}

\usepackage{geometry}
\usepackage{subcaption}
\usepackage{graphicx}
\usepackage{mathtools, makecell}

\usepackage{hyperref}
\usepackage{cleveref} 

\newcommand{\scriptR}{\mathcal{R}}

\usepackage{setspace}

\begin{document}

\pagenumbering{gobble}

\title{Emergent Universe from an Unstable de Sitter Phase}

\maketitle

\begin{center}
    
\author{Molly Burkmar${{^{a,1}}}$ (corresponding author) and Marco Bruni${{^{a,b,2}}}$}

\vspace{0.5cm}

${{^a}}$Institute of Cosmology {\rm \&} Gravitation, University of Portsmouth, Dennis Sciama Building, Burnaby Road, Portsmouth, PO1 3FX, United Kingdom\\
${{^b}}$INFN Sezione di Trieste, Via Valerio 2, 34127 Trieste, Italy

\vspace{0.5cm}

$^{1}$molly.burkmar@port.ac.uk

$^{2}$marco.bruni@port.ac.uk

\vspace{0.5cm}

Essay written for the Gravity Research Foundation 2024 Awards for Essays on Gravitation.

\vspace{0.5cm}
\end{center}
\begin{abstract}
    
In the Emergent scenario, the Universe should evolve  from a non-singular state replacing the typical singularity of General Relativity,  for any  initial condition. For the scalar field  model in \cite{Ellis2003} we show that only a set of measure zero of trajectories leads to emergence, either from a static state (an Einstein model), or from a de Sitter state.

Assuming a scenario based on CDM interacting with a  Dark Energy fluid, we show that in general flat and open models expand from a non-singular unstable de Sitter  state at high energies; for some closed models this state is a transition phase with a bounce, other closed models are cyclic. A subset of these models are qualitatively in agreement with the observable Universe, accelerating at high energies, going through a matter-dominated decelerated era, then accelerating toward a de Sitter phase.

\end{abstract}
\newpage
\pagenumbering{arabic}

\section{Introduction}

The Standard Model of Cosmology, a flat $\Lambda$CDM model with inflation, is based on 
General Relativity (GR) and 
provides a successful framework to describe the evolution of the Universe. However, the presence of singularities pose a problem and these are currently interpreted as points in space-time where GR breaks down \cite{Joshi1993,Hawking1979,Gibbons2003,Ashtekar2015}. Observations are consistent with a flat universe \cite{Planck2018I, Planck2018VI}, however they do not rule out open or closed spatial curvature \cite{DiValentino:2020srs}. In particular, a closed universe is appealing as it avoids the problem of having an infinite universe \cite{EllisInfinity}. The Emergent Universe has been widely studied as an alternative to an initial singularity, 
with models  built with fine-tuned initial conditions on the assumption that the past repellor must be a static  state represented by an Einstein model with positive curvature
\cite{Ellis2003,Ellis2004,Mukherjee2006,Verlinde2017,Ellis2016,Parisi2007,Mulryne2005,Bonanno2017,Gionti2017, Sengupta2023, Barrow2003}. Thus the problem with this scenario is that very special initial conditions are required, as the Einstein model is typically represented by an unstable saddle or a centre in the phase space of the relevant Ordinary Differential Equations (ODE), and so it is not a generic past repellor, as in the classical model by Eddington \cite{Eddington1930}.

In this essay we contend that in the Emergent scenario, the Universe should evolve from a non-singular state replacing the typical singularity of GR for any initial condition. In this light, we first consider whether models in \cite{Ellis2003}, based on a scalar field, are still non-singular for general initial conditions. Similarly to the analysis in \cite{Amendola1990} and \cite{Belinskii1985}, we complete a dynamical systems analysis of the Emergent scenario in \cite{Ellis2003}. We show that only a set of measure zero of trajectories emerge from the static Einstein model. We also find emergence from a non-singular de Sitter phase, however like with emergence from the Einstein state, only a set of measure zero of trajectories emerge from this phase. For trajectories not emerging from these states, we find that the past repellor is a singularity where the scalar field is kinetic dominated.\footnote{The analysis can be found at: \url{https://github.com/MollyBurkmar1/Emerging_Universe.git}}.

We then consider a fluid-based model to show an example where emergence from a non-singular state is generic during expansion, regardless of initial conditions. Assuming a scenario based on Cold Dark Matter (CDM, represented by a dust fluid) interacting with a  Dark Energy fluid with a nonlinear equation of state, we show that all but a set of measure zero of trajectories in phase space are either cyclic or expand from an unstable de Sitter state, a classical vacuum at high energies. This represents the asymptotic past of flat and open models, and a transition phase with a bounce for closed models. Independently from the initial conditions, all models are accelerating at high energies. A subset of models are qualitatively in agreement with the observed universe, emerging from a non-singular state, going through a matter-dominated decelerated era, then accelerating toward the future de Sitter phase.

\section{Scalar-Field-Based Emergent Universe}

We consider a Friedmann-Lema{\^i}tre-Robertson-Walker (FLRW) universe, containing a scalar field $\phi$ with the potential

\begin{equation}
    V = V_0 \left(e^\phi - 1\right)^2 \,,
    \label{eqn:potential}
\end{equation}

\noindent where $V_0$ is the asymptotic value of the potential  at $\phi \rightarrow -\infty$. Assuming $8\pi G = c = 1$, the Klein-Gordon and Raychaudhuri equations are

\begin{equation}
    \Ddot{\phi} = -3H\Dot{\phi} - 2V_0e^{\phi}(e^{\phi}-1) \,,
    \label{eqn:KG_dimensions}
\end{equation}

\begin{equation}
    \Dot{H} = -H^2 - \frac{1}{3}\left[\Dot{\phi}^2 - V_0(e^\phi - 1)^2 \right] \,,
\end{equation}

\noindent where $H=\dot{a}/a$ is the Hubble expansion scalar ($a$ is the scale factor) and over-dots are derivatives with respect to time. 
The dynamics is complemented by the usual Friedmann first integral (the Hamiltonian constraint) 
\begin{equation}
    H^2= -\frac{k}{a^2} +\frac{1}{3} \left[\frac{\dot{\phi}^2}{2} +V(\phi)\right].
    \label{eqn:Friedmann}
\end{equation}
Setting $\psi= \Dot{\phi}/\sqrt{V_0}$ defines a 3-dimensional  set of ODE.  
 To analyse this dynamical system we define dimensionless variables ($\phi$ is dimensionless by definition) and their  compactified version,

\begin{equation}
    \eta = \sqrt{V_0}t, ~~~ \Phi = \frac{\phi}{\sqrt{1+\phi^2}}, ~~~ \psi = \frac{\Dot{\phi}}{\sqrt{V_0}}, ~~~ \Psi = \frac{\psi}{\sqrt{1+\psi^2}}, ~~~ y = \frac{H}{\sqrt{V_0}}, ~~~ Y = \frac{y}{\sqrt{1+y^2}} \,,
    \label{eqn:dimensionless}
\end{equation}
in order to study the behaviour of the system at infinity.
Here, $\phi \rightarrow \pm \infty$ corresponds to $\Phi = \pm 1$, $\psi \rightarrow \pm \infty$ to $\Psi = \pm 1$ and $y \rightarrow \pm \infty$ to $Y = \pm 1$.

First, we consider the phase space of spatially flat models. The Friedmann equation \eqref{eqn:Friedmann} with $k = 0$ in our compactified variables becomes

\begin{equation}
    \frac{Y}{\sqrt{1-Y^2}} = \pm \sqrt{\frac{1}{3}\left(e^\frac{\Phi^2}{\sqrt{1-\Phi^2}}-1\right)^2 + \frac{\Psi^2}{6\left(1-\Psi^2\right)}}\,,
    \label{eqn:Friedmann_flat}
\end{equation}

\noindent which we substitute into the equation for $\Psi$, and numerically solve alongside the equation for $\Phi$. Here, $Y > 0$ corresponds  to expansion and $Y < 0$ to contraction. 

The fixed points of the flat sub-manifold with their stability are summarised in Table \ref{tab:Flat_FPs}. To understand the type of fixed points present at $\Phi = \pm 1$ and $\Psi = \pm 1$, we need to understand what happens to the potential \eqref{eqn:potential} and kinetic energy $K = \psi^2/2$ at these points. When $\Psi = \pm 1$, $K \rightarrow \infty$. For the fixed points where $\Phi = -1$, the potential is finite: $V = V_0$. In this case the fixed points are kinetic dominated and so are stiff fluid singularities. At $\Phi = +1$, both the potential energy and kinetic energy become infinite, and the fixed points cannot be linearised. We find that the kinetic energy grows more quickly than the potential, and so these fixed points represent singularities.

Only the stability of the de Sitter fixed points $dS_{1\pm}$ can be found from linearisation of the system; for the other fixed points there are infinities in the Jacobian. We can find their stability through plots of the phase space, however for repellor and attractor fixed points it is helpful to define Liapunov functions \cite{Arrowsmith1982,Arrowsmith1992} in order to obtain more rigorous results. For the Minkowski fixed point at $\phi = 0$, we define the Liapunov function $L$ as the energy density of the scalar field, where we approximate $e^\phi \simeq 1 + \phi$ such that $V \simeq \phi^2$ around the origin:

\begin{equation}
    L = \frac{\Phi^2}{1-\Phi^2} + \frac{\Psi^2}{2\left(1-\Psi^2\right)}.
    \label{eqn:Liapunov_Flat_Minkowski}
\end{equation}

\noindent We then take the first derivative of $L$ to be able to determine the stability of the fixed point. If $L' < 0$ at least in a disc around the fixed point, then it is is an attractor, and if $L' > 0$ then the fixed point is a repellor. Taking the first derivative of $L$, we find

\begin{equation}
    L' = - \frac{3\Psi^2 Y}{(1-\Psi^2)\sqrt{1-Y^2}}.
\end{equation}

Taking $Y > 0$ ($Y < 0)$ for the expanding (contracting) case, we find $L' < 0$ ($L' > 0)$ for the whole phase space, therefore the Minkowski fixed point is a global attractor (repellor) in the expanding (contracting) phase space. For the past (future) singularity, we define the Liapunov function as a circle centered around the fixed point:

\begin{equation}
    L = \left(1+\Phi\right)^2 + \left(\Psi \mp 1\right)^2 \,.
    \label{eqn:L_singularities}
\end{equation}

\noindent We fix a radius of $L_* = r^2 = 0.5^2$ so that we only study the behaviour of trajectories around the singularities. The first derivative of $L$ gives

\begin{equation}
    \L' = \frac{2\Psi\left(1+\Phi\right)\left(1-\Phi^2\right)^{3/2}}{\sqrt{1-\Psi^2}} + 2\left(\Psi \mp 1\right)\left[-2\left(1-\Psi^2\right)^{3/2}e^{\frac{\Phi}{\sqrt{1-\Phi^2}}}\left(e^{\frac{\Phi}{\sqrt{1-\Phi^2}}} - 1\right) - \frac{3Y\Psi\left(1-\Psi^2\right)}{\sqrt{1-Y^2}}\right]\,.
    \label{LDot_Flat_S}
\end{equation}

\noindent For the expanding (contracting) case, we find $L' > 0$ ($L'< 0$) for $L < L_*$, therefore the past (future) singularity is a repellor (an attractor).

\begin{table}[!ht]
    \centering
    \begin{tabular}{|c|c|c|c|c|c|}
    \hline
        Fixed point & $\Phi$ & $\Psi$ & $Y$ & Stability in 2-D & Stability in 3-D \\
         \hline
         $M_\pm$ & 0 & 0 & 0 & Spiral attractor (+) / Spiral repellor (-) & Spiral saddle \\
         $dS_{1\pm}$ & $-1$ & 0 & $\pm 1/2 $ & Saddle ($\pm$) &  Saddle ($\pm$) \\
         $S_{1\pm}$ & $-1$ & $1$ & $\pm 1$ & Repellor (+)/ Saddle (-) & Repellor (+)/ Saddle (-)\\
         $S_{2\pm}$ & $-1$ & $-1$ & $\pm 1$ & Saddle (+)/ Attractor (-) & Saddle (+)/ Attractor (-)\\
         $S_{3\pm}$ & $1$ & $1$ & $\pm 1$ & Saddle ($\pm$) & Saddle ($\pm$)\\
         $S_{4\pm}$ & $1$ & $-1$ & $\pm 1$ & Saddle ($\pm$) & Saddle ($\pm$)\\
         \hline
    \end{tabular}
    \caption{Fixed points for the expanding (+) and contracting (-) flat models in the 2-D sub-manifolds in Fig. \ref{flat_submanifold}, and the full 3-D phase space in Fig. \ref{fig:full_phase_spaces}. $M$ denotes the Minkowski fixed point, $dS$ the de Sitter fixed points and $S$ the fixed points representing singularities.}
    \label{tab:Flat_FPs}
\end{table}  

\begin{figure}
\begin{subfigure}[h]{0.47\linewidth}
\includegraphics[width=\linewidth]{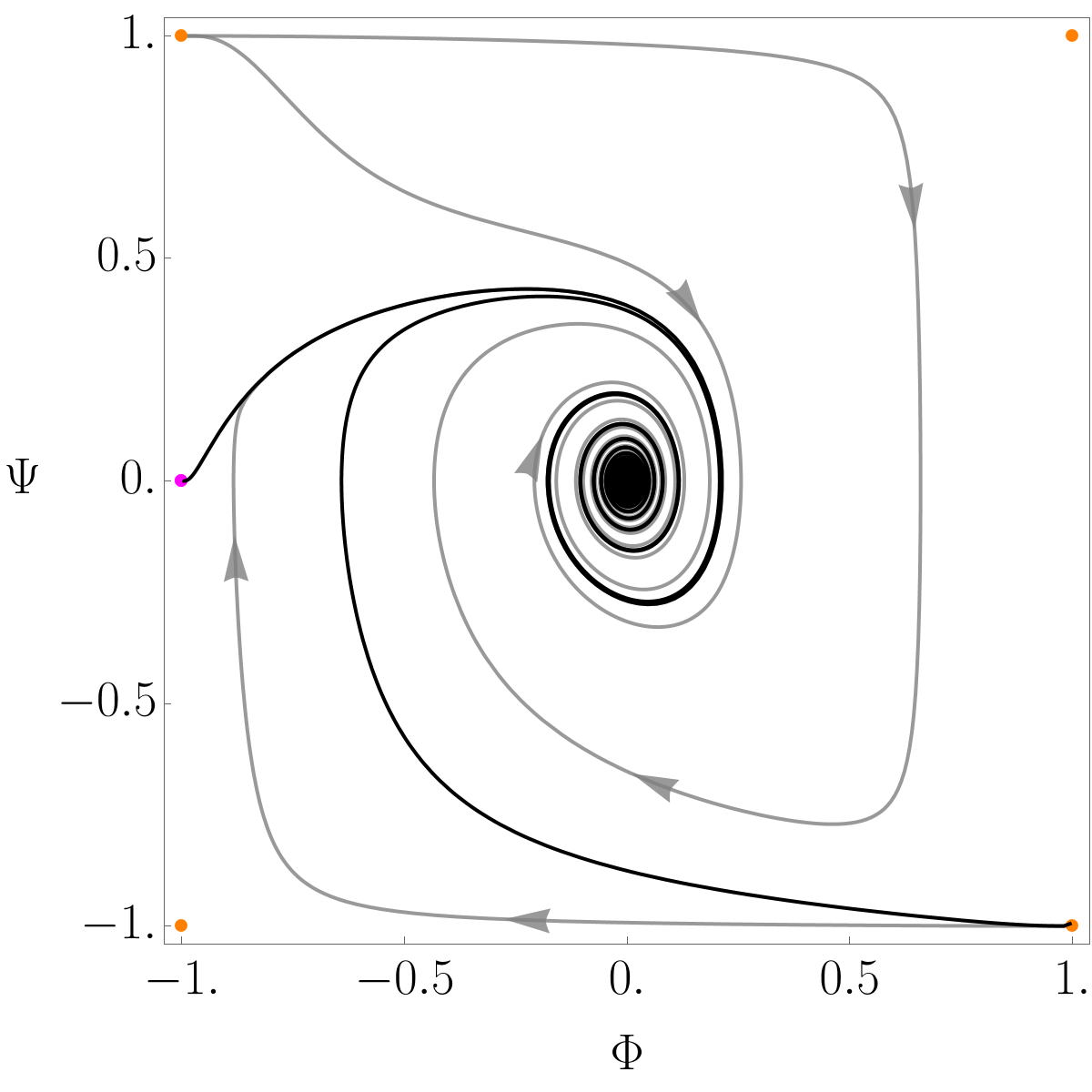}
\caption{Expansion ($Y > 0$).}
\label{flat_submanifold_exp}
\end{subfigure}
\hfill
\begin{subfigure}[h]{0.47\linewidth}
\includegraphics[width=\linewidth]{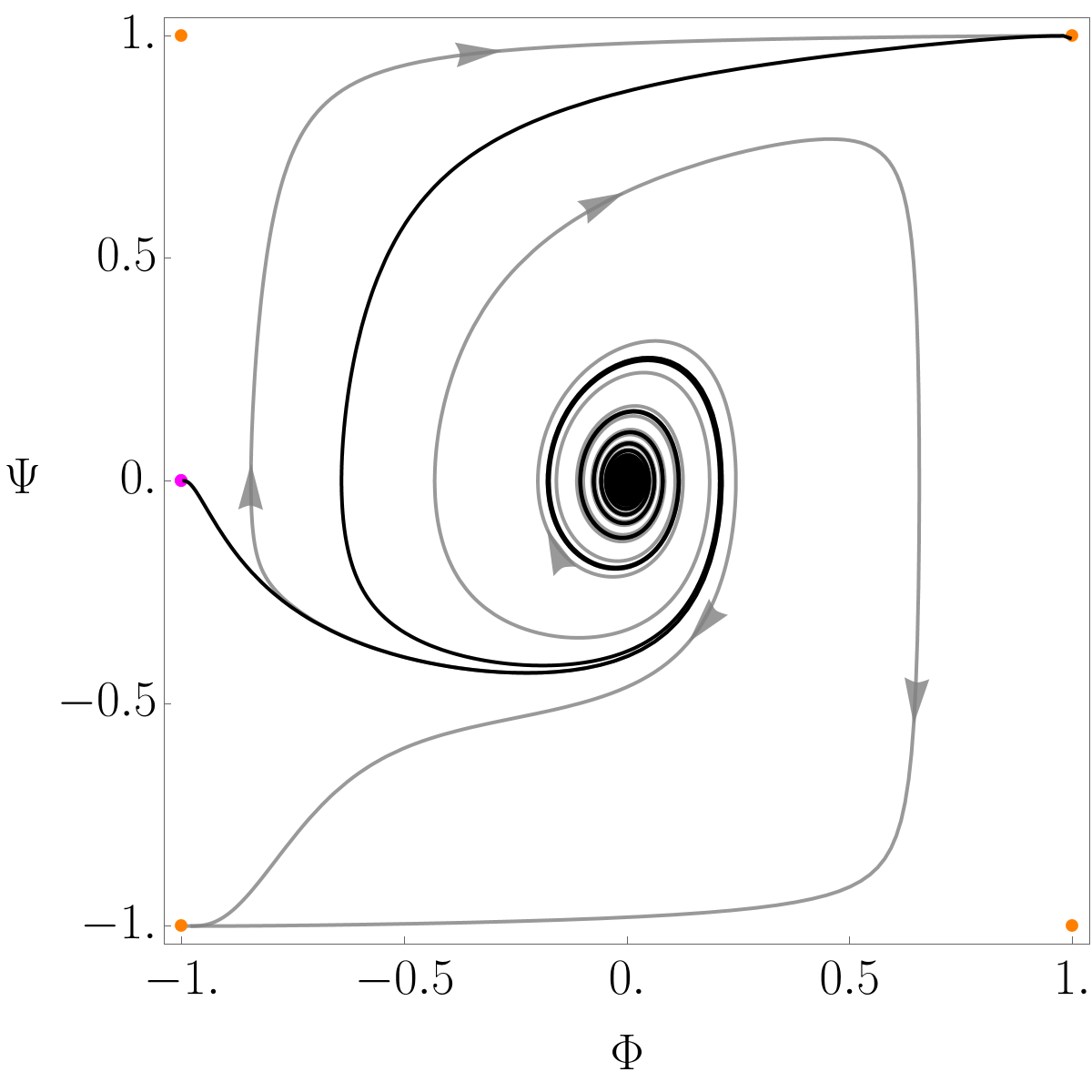}
\caption{Contraction ($Y < 0$).}
\label{flat_submanifold_con}
\end{subfigure}
\caption{The flat sub-manifolds for the system, where we substitute $Y > 0$ for the expanding case in (a) and $Y < 0$ for the contracting case in (b). The Minkowski fixed point is shown in black, the de Sitter fixed points in magenta and the singularities in orange. The two black curves are separatrices that separate where the trajectories expand from (contract to). Trajectories between the two separatrices expand from (contract to) the singularity $S_{4+}$ ($S_{3-}$). Otherwise, trajectories outside these separatrices expand from (contract to) $S_{1+}$ ($S_{2-}$).}
\label{flat_submanifold}
\end{figure}

The phase space for the flat sub-manifold is shown in Fig. \ref{flat_submanifold}, where Fig. \ref{flat_submanifold_exp} shows the expanding case, and Fig. \ref{flat_submanifold_con} the contracting case. In the expanding case, trajectories between the two separatrices expand from $S_{4+}$, whereas trajectories outside the two separatrices expand from $S_{1+}$. In both cases trajectories are kinetic dominated in the past. The past repellor $S_{1+}$ represents the scalar field starting on the flat part of the potential and rolling towards the minimum, and the fixed point $S_{4+}$ represents the field starting from the exponential part of the potential and rolling towards the minimum. There is also a special trajectory in the phase space, which emerges from the de Sitter fixed point $dS_{1+}$ and expands towards the Minkowski fixed point $M$. In this case, the field is initially on the asymptotic part of the potential, and rolls toward the minimum where it oscillates.

Fig. \ref{flat_submanifold_con} shows the contracting case. Here, trajectories contract from the Minkowski fixed point $M$ in all cases, and in general contract towards a singularity at $S_{2-}$ or $S_{3-}$. Trajectories between the two separatrices contract towards $S_{3-}$, and trajectories outside the separatrices contract towards the singularity $S_{2-}$. In both cases trajectories become kinetic dominated at the singularity. There is also the possibility that a trajectory contracts towards the de Sitter fixed point $dS_{1-}$, although this requires a set of measure zero initial conditions.

It is clear there are flat models for which trajectories are non-singular and can be past or future asymptotic to a de Sitter state $dS_{1\pm}$. However, these fixed points have saddle stability and are a set of measure zero in the phase space. For all other trajectories the past repellor is either $S_{1+}$ or $S_{4+}$.

\begin{figure}[!ht]
\begin{center}
\includegraphics[width=100mm]{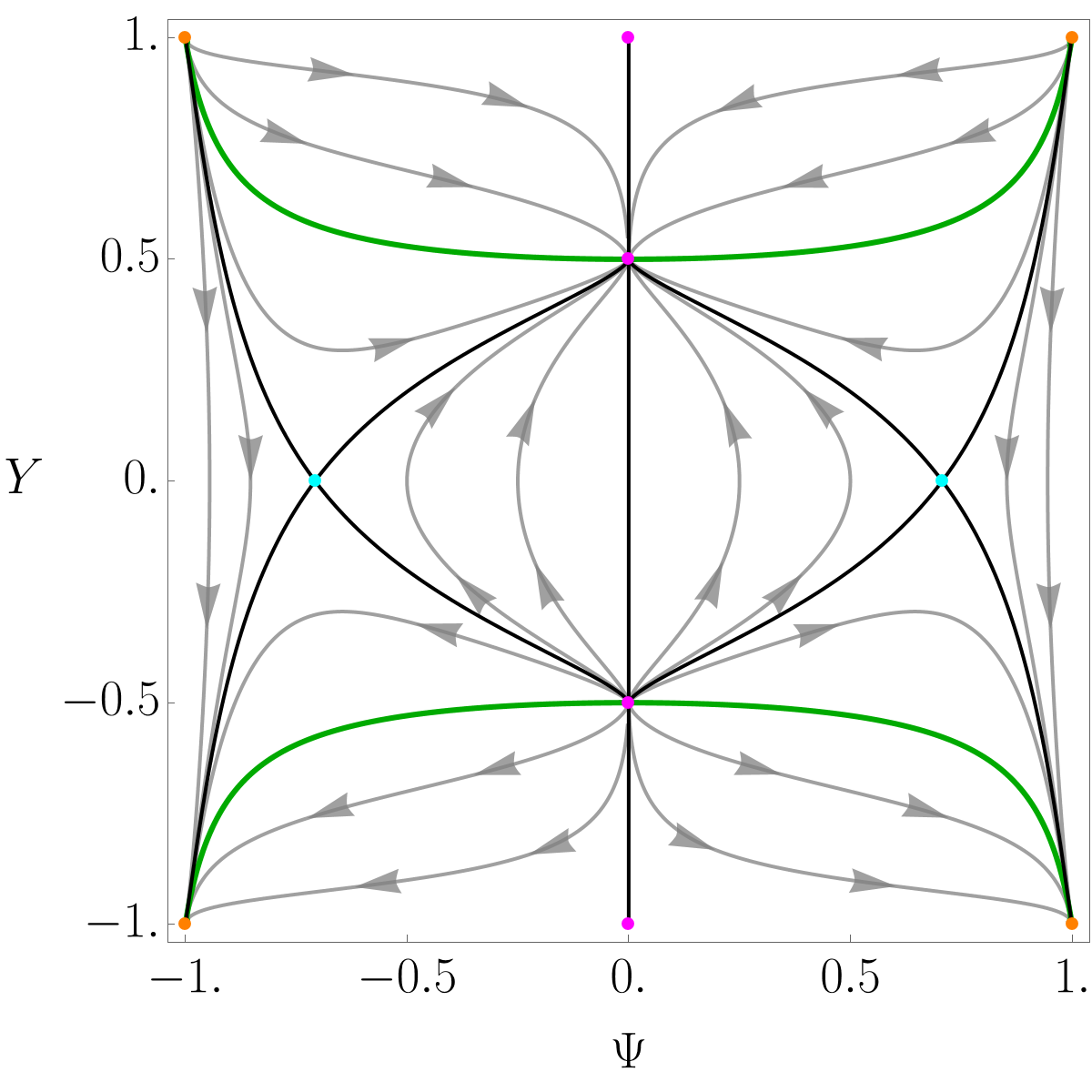}
\end{center}
\caption{The $\Phi = -1$ submanifold. The thick green curve shows the Flat Friedmann Separatrix (FFS) separating the open models outside the FFS from the closed models between the two green curves. The thick black curve is the Closed Friedmann Separatrix (CFS) which seperates different types of closed models in the phase space. The Einstein fixed points are shown in cyan, the de Sitter fixed points in magenta and the singularities in orange.}
\label{fig:Phi_Minus_1}
\end{figure}

Going back to the 3-D dynamics, we first consider the $\Phi = -1$ sub-manifold, where the potential \eqref{eqn:potential} reduces to $V = V_0$. The fixed points of the $\Phi = -1$ sub-manifold are shown in Table \ref{tab:Phi_Submanifold_FPs}, and the phase space is shown in Fig. \ref{fig:Phi_Minus_1}. The Einstein points $E_{\pm}$ are both unstable saddles, with only two separatrix trajectories emerging from each. Therefore, these are not good candidates for emerging models as these are a set of measure zero for the whole phase space. In this sub-manifold, the past repellors for initially expanding models are the singularities $S_{1+}$ and $S_{2+}$, and the de Sitter fixed point $dS_{1-}$ is the past repellor for models which initially contract. Trajectories emerging from the non-singular de Sitter state $dS_{1-}$ either collapse, bounce and expand towards another de Sitter state $dS_{1+}$, or collapse to a singularity ($S_{1-}$ or $S_{2-}$).

\begin{table}[!ht]
    \centering
    \begin{tabular}{|c|c|c|c|c|}
        \hline
        Fixed point & $\Psi$ & $Y$ & Stability in 2-D & Stability in 3-D \\
        \hline
         $E_{\pm}$ & $\pm\frac{1}{\sqrt{2}}$ & 0 & Saddle ($\pm$) & Saddle ($\pm$) \\
         $dS_{1\pm}$ & 0 & $\pm\frac{1}{2}$ & Attractor (+)/ Repellor (-) & Saddle ($\pm$) \\
         $dS_{2\pm}$ & 0 & $\pm 1$ & Saddle ($\pm$) & Saddle ($\pm$) \\
         $S_{1\pm}$ & 1 & $\pm 1$ & Repellor (+)/ Attractor (-) & Repellor (+)/ Saddle (-)\\
         $S_{2\pm}$ & -1 & $\pm1$ & Repellor (+)/ Attractor (-) & Saddle (+)/ Attractor (-) \\
        \hline
    \end{tabular}
    \caption{Fixed points with their stability for the 2-D $\Phi = -1$ sub-manifold in Fig. \ref{fig:Phi_Minus_1} and the full 3-D phase space in Fig. \ref{fig:full_phase_spaces}. $E$ denotes the Einstein fixed points, $dS$ the de Sitter fixed points and $S$ the singularities. Note that $dS_{1\pm}$ represent flat de Sitter models, and $dS_{2\pm}$ represent open de Sitter models with coordinate singularities.}
    \label{tab:Phi_Submanifold_FPs}
\end{table}

\begin{figure}[!ht]
\begin{subfigure}[h]{0.49\linewidth}
\includegraphics[width=\linewidth]{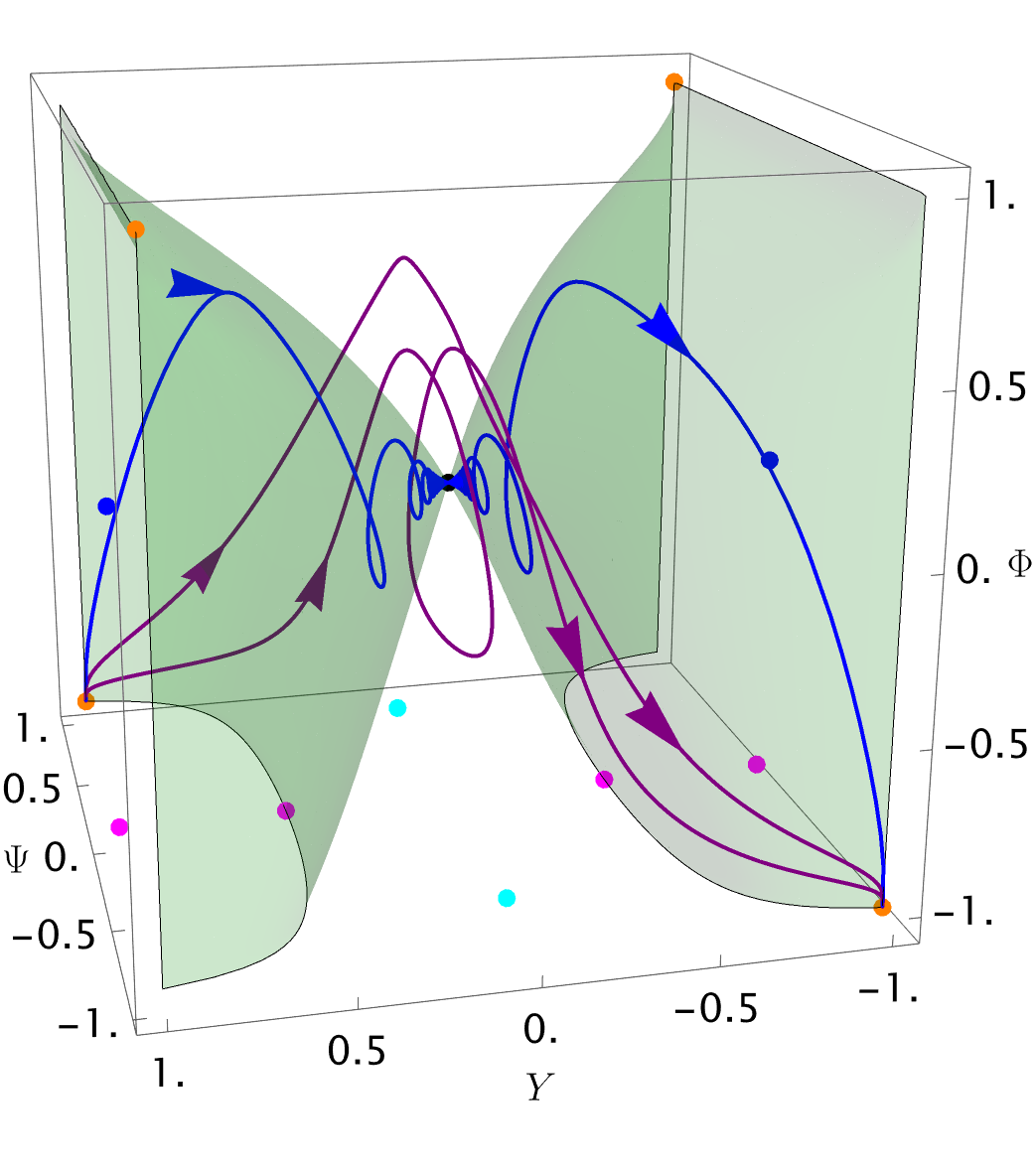}
\caption{Singular trajectories.}
\label{fig:full_open_closed}
\end{subfigure}
\hfill
\begin{subfigure}[h]{0.49\linewidth}
\includegraphics[width=\linewidth]{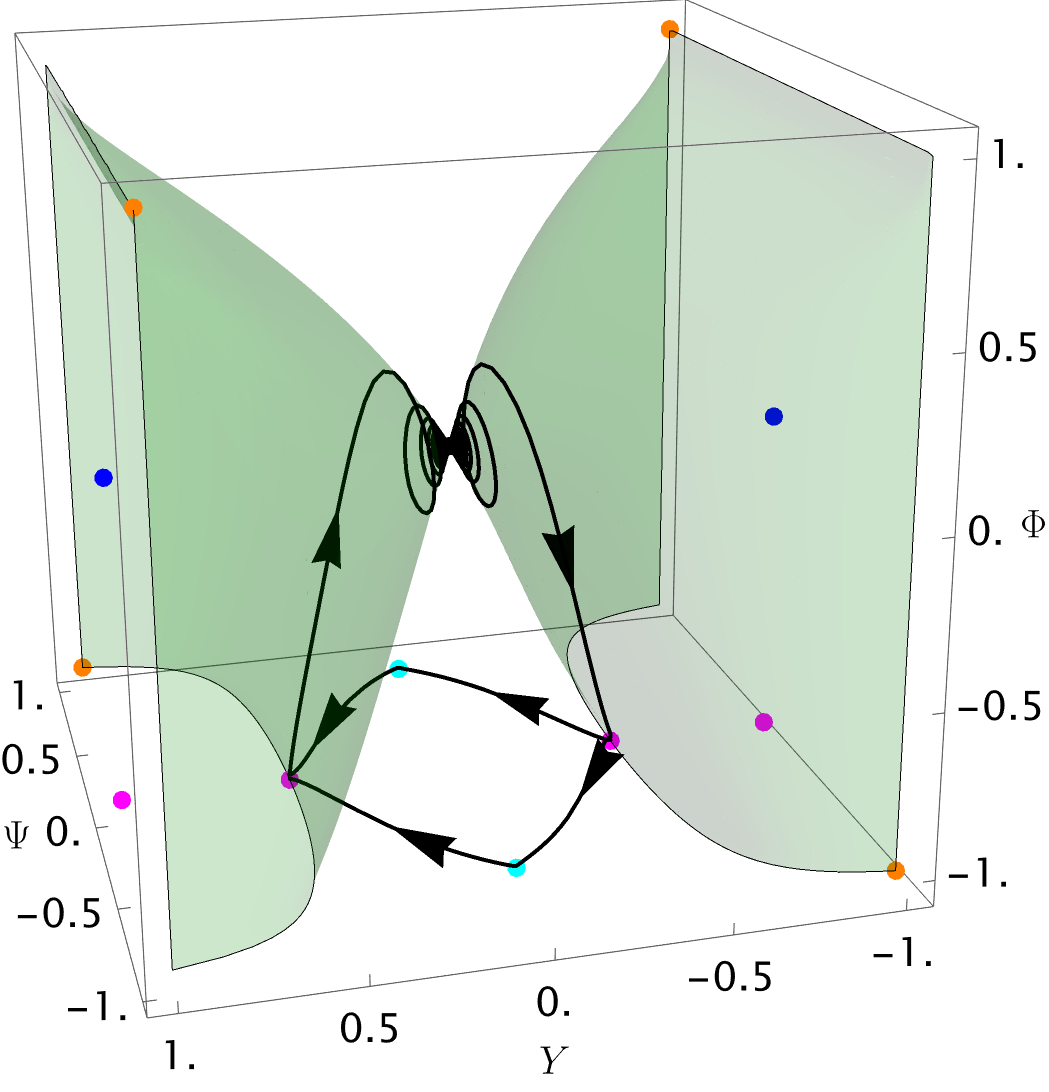}
\caption{Non-singular trajectories.}
\label{fig:full_non_singular}
\end{subfigure}
\caption{The full phase space for the system, where we have plotted examples of open (blue) and closed (purple) trajectories, as well as separatrices (black). The green surface shows the Flat Friedmann Separatrix. The Minkowski fixed point is shown in black, the Einstein fixed points in cyan, the de Sitter fixed points in magenta, the Milne fixed points in blue and the singularities in orange.}
\label{fig:full_phase_spaces}
\end{figure}

The full phase space for the 3-D system is shown in Fig. \ref{fig:full_phase_spaces}, where Fig. \ref{fig:full_open_closed} shows examples of open (blue) and closed (purple) trajectories coming from and going to a singularity, and Fig. \ref{fig:full_non_singular} shows examples of non-singular trajectories. The fixed points that have not already been considered in the sub-manifolds are shown in Table \ref{tab:Full_FPs}. In Fig. \ref{fig:full_open_closed}, there are two examples of closed models: one simply expands from a singularity, reaches a turn-around on the surface $Y = 0$ and then collapses to a singularity. The other expands and oscillates around the Minkowski fixed point (at the minimum of the potential), where it contracts, bounces and expands, before turning around on the $Y = 0$ surface and collapsing to the singularity. Whether there is a sufficient inflationary phase after the bounce remains to be seen. The expanding open model evolves from a singularity, and oscillates around the minimum of the potential, and the contracting open model oscillates around the Minkowski fixed point before collapsing to a singularity. 

Fig. \ref{fig:full_non_singular} shows examples of non-singular models in the phase space, which emerge from the Einstein fixed points $E_{\pm}$ (cyan) and the de Sitter fixed point $dS_{1+}$ (magenta). The trajectories that emerge from these fixed points require extremely fine-tuned initial conditions: they form a set of measure zero in phase space.Trajectories can also emerge from $dS_{2+}$ and $Mi_{+}$, but these are also a set of measure zero in the phase space. For all other trajectories the past repellor is either $S_{1+}$ or $S_{4+}$. The singularity $S_{4+}$ is a 3-D saddle, repelling the aforementioned subset and attracting in another sub-manifold, and the $S_{1+}$ point represents a stiff fluid singularity where the scalar field is kinetic dominated. Trajectories emerging from $S_{4+}$ require some fine-tuning, but regardless most trajectories expand from a singularity.

\begin{table}[]
    \centering
    \begin{tabular}{|c|c|c|c|c|}
        \hline
        Fixed point & $\Phi$ & $\Psi$ & $Y$ & Stability \\
        \hline
         $Mi_{\pm}$ & 0 & 0 & $\pm1$ & Saddle ($\pm$) \\
        \hline
    \end{tabular}
    \caption{The fixed points of the full system not already shown in the flat and $\Phi = -1$ sub-manifolds. Note that the $Mi_{\pm}$ fixed points represent Milne models (Minkowski in open coordinates), which have coordinate singularities.}
    \label{tab:Full_FPs}
\end{table}


\section{Emergence from Interacting Dark Energy and Dark Matter}

In this essay, we contend that it is desirable for all trajectories to emerge from a non-singular state in the Emerging Universe scenario, regardless of initial conditions. In the following, we provide an example of a model with dark energy with a non-linear equation of state \cite{Burkmar2023} interacting with a dust fluid \cite{Burkmar2024}, where the generic trajectory either represents cyclic models, or models that emerge from an unstable de Sitter phase during expansion. For flat and open models this is the asymptotic past repellor, whilst for a subset of closed models the de Sitter phase represents a transition through a bounce from contraction to expansion. 

In this scenario, the conservation equations for the dark matter and dark energy are

\begin{equation}
    \dot{\rho}_m=-3H\rho_m + \frac{H\rho_x\rho_m}{\rho_{i}}\,,
    \label{rhomdot_interacting}
\end{equation}

\begin{equation}
    \dot{\rho_x}=-3H(\rho_x-\rho_\Lambda)\left(1 + w_x + \epsilon\frac{\rho_x}{\rho_*}\right)-\frac{H\rho_x\rho_m}{\rho_i}\,.
    \label{rhoxdot_interacting}
\end{equation}

\noindent where $\rho_m$ is the dark matter energy density and $\rho_x$ is the dark energy density. The dark energy has a non-linear equation of state, with $w_x$ defining its linear part, $\rho_*$ the characteristic scale of the non-linear part, and $\rho_\Lambda$ representing its low energy attractor. $\epsilon$ is a free dimensionless parameter that fixes the sign and the strength of the quadratic term and the energy scale $\rho_i$ characterises the non-linear interaction. To close the system, we also require the Raychaudhuri equation to describe the evolution of the expansion scalar $H$,

\begin{equation}
    \dot{H}=-H^2-\frac{1}{6}\left[\rho_m + \rho_x(1 + 3w_x - 3\epsilon\frac{\rho_\Lambda}{\rho_*}) \\
    - 3\rho_\Lambda(1 + w_x) + 3\epsilon\frac{\rho_x^2}{\rho_*}\right]\,.
    \label{Hdot_interacting}
\end{equation}

\noindent To analyse the dynamical system consisting of \eqref{rhomdot_interacting}, \eqref{rhoxdot_interacting} and \eqref{Hdot_interacting}, we first define dimensionless variables,

\begin{equation}
    x=\frac{\rho_x}{\rho_*}, ~~ y=\frac{H}{\sqrt{\rho_*}}, ~~ z=\frac{\rho_m}{\rho_*}, ~~ \scriptR = \frac{\rho_\Lambda}{\rho_*}, ~~ q = \frac{\rho_*}{\rho_i}, ~~ \eta=\sqrt{\rho_*}t\,,
    \label{eqn:normalisation}
\end{equation}

\noindent as well as compactified variables,

\begin{equation}
    X = \frac{x}{1+x}, ~~ Y = \frac{y}{\sqrt{1+y^2}}, ~~ Z = \frac{z}{1+z} \,.
\end{equation}

\noindent in order to see the behaviour of the system at infinity. For this system, $Y = \pm 1$ corresponds to $H \rightarrow \pm \infty$, and $X = Z = 1$ corresponds to $\rho_x, \rho_m \rightarrow +\infty$.

\begin{figure}[!ht]
\begin{subfigure}[h]{0.58\linewidth}
\includegraphics[width=\linewidth]{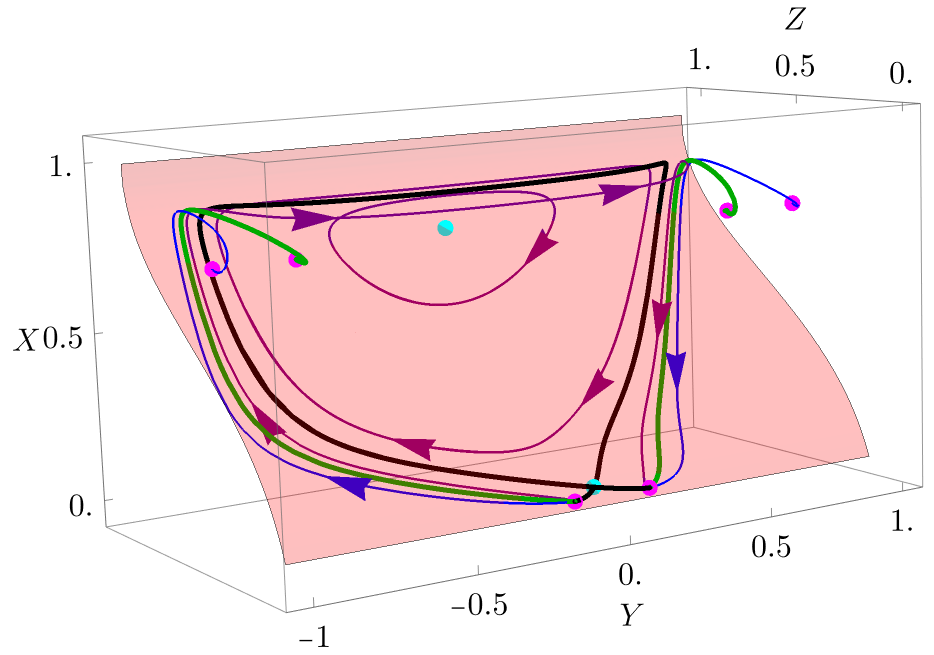}
\caption{}
\label{fig:Int_DE_Front}
\end{subfigure}
\hfill
\begin{subfigure}[h]{0.42\linewidth}
\includegraphics[width=\linewidth]{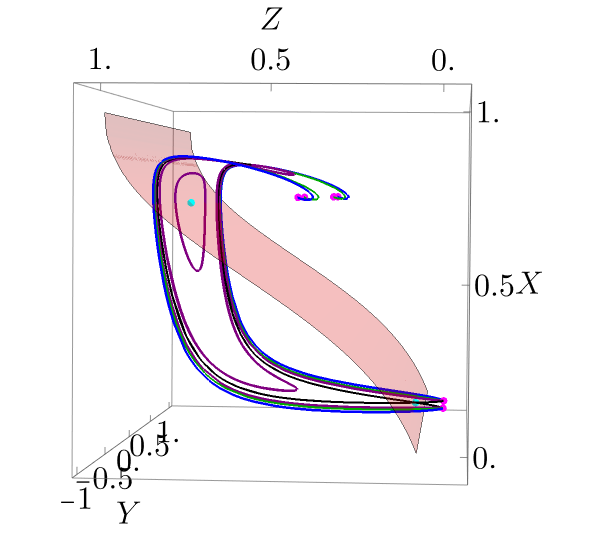}
\caption{}
\label{fig:Int_DE_Side}
\end{subfigure}
\caption{The full phase space for the interacting cold dark matter and dark energy model, where the parameters are set to $\epsilon = -0.25$, $q = 1$, $\scriptR = 0.05$ and $w_{x} = -0.5$. The reddish surface shows where the acceleration is zero, with trajectories accelerating above this surface and decelerating below it. The green trajectories show the flat models and the black curves show the Closed Friedmann Separatrix (CFS). Open models are shown in blue, and closed models are shown in purple. Some closed models are cyclic. All other trajectories emerge from a non-singular state, generically from a de Sitter fixed point (magenta) and from an Einstein fixed point (cyan) for a special initial condition where $Y=0$. We only show the fixed points that are relevant to the dynamics; the full list of fixed points can be seen in Table \ref{tab:Int_DE_FPs}.}
\label{fig:Int_DE}
\end{figure}

An example of the full phase space for a subset of models that expand towards a future de Sitter state is shown in Fig. \ref{fig:Int_DE} with the fixed points shown in Table \ref{tab:Int_DE_FPs}. We set the parameter values to $\epsilon = -0.25$, $q = 1$, $\scriptR = 0.05$\footnote{Note that we set $\scriptR$ such that the phase space is readable. In reality we would expect $10^{-120} < \scriptR < 10^{-60}$.} and $w_{x} = -0.5$, which defines a surface in the phase space. The reddish surface shows where the acceleration is zero, with models accelerating when trajectories are above this surface, and decelerating when they are below it. The Flat Friedmann Separatrix would be a surface in the phase space, however for the sake of clarity we just show the flat trajectories in green for the parameter values we set. There is a trajectory, not shown in the figure, which connects the flat fixed points $dS_{3+}$ and $dS_{3-}$. This is a high energy de Sitter trajectory, representing a closed de Sitter model.

In this example, all trajectories are non-singular. Generically they either emerge from a de Sitter phase or are cyclic models; a separatrix represents emergence from an Einstein state. Open (blue) and flat models emerge from a de Sitter fixed point (magenta) and expand (contract) to another de Sitter state. The Closed Friedmann Separatrix (CFS) passing through the saddle Einstein point is shown by the thick black curve. Closed models (purple) outside the CFS emerge from a contracting phase asymptotic in the past to the de Sitter fixed point $dS_{1-}$, then bounce at high energy before expanding to $dS_{1+}$. Trajectories inside the CFS either bounce once, contracting from $dS_{1-}$ and expanding to $dS_{1+}$, or are cyclic around the Einstein fixed point $E_2$, and repeatedly contract, bounce and expand. An emergent case which expands from the Einstein fixed point $E_1$ also exists along the CFS. 

In this example, two Einstein fixed points (cyan) exist in the phase space, which coincide with where the acceleration is zero. This means in this example, open, flat and closed trajectories outside of the CFS all evolve with an early- and late-time acceleration, connected by a decelerated period. These are the cases of interest as they qualitatively match the observed Universe.

\begin{table}[]
    \centering
    \begin{tabular}{|c|c|c|c|c|}
    \hline
        Fixed point & $X$ & $Y$ & $Z$ & Stability \\
         \hline
          $dS_{1\pm}$ & 0.048 & $\pm$0.13 & 0 & Attractor (+)/ Repellor (-) \\
          $dS_{2\pm}$ & 0.67 & $\pm$ 0.63 & 0 & Saddle ($\pm$) \\
          $dS_{3\pm}$ & 0.75 & $\pm$ 0.74 & 0.42 & Spiral saddle ($\pm$) \\
          $dS_{4\pm}$ & 0.75 & $\pm$ 1 & 0.42 & Spiral repellor (+)/ Spiral attractor (-) \\
          $E_1$ & 0.052 & 0 & 0.094 & Saddle  \\
          $E_2$ & 0.73 & 0 & 0.87 & Centre \\
         \hline
    \end{tabular}
    \caption{Fixed points for the fluid model, where dust is interacting with a dark energy with a non-linear equation of state. Note that $dS_{4\pm}$ represent open de Sitter models with coordinate singularities. The parameters are set to $\epsilon = -0.25$, $q = 1$, $\scriptR = 0.05$ and $w_{x} = -0.5$.}
    \label{tab:Int_DE_FPs}
\end{table}

\section{Conclusions}
In this essay, we have carried out a dynamic systems analysis of the Emergent Universe scenario in \cite{Ellis2003}, which consists of a scalar field with an asymmetric potential with a plateau for negative $\phi$. Originally, this scenario was considered with specific initial conditions such that the universe emerged from a static Einstein state. The aim of this essay was to understand whether the past repellor was generically non-singular for any initial conditions. 
The de Sitter model is normally associated to a cosmological constant associated with an asymptotically stable future attractor \cite{Wald:1983ky,Bruni:2001pc}, but it can also represent a past repelling fixed point \cite{Ananda2006,Burkmar2023}.
We found trajectories which emerge from a de Sitter fixed point, however like with emergence from the Einstein fixed point, they form a set of measure zero in phase space. All other trajectories expand from a singularity. We remark that simply with the exchange $\phi\rightarrow-\phi$ the potential \eqref{eqn:potential} for the model in \cite{Ellis2003} becomes the potential for the Einstein-frame version \cite{Maeda1988} of Starobinsky inflation \cite{Starobinsky1980}, hence our analysis can  easily be mapped into one for that model. Models of this type are currently the most favored by experimental results \cite{Planck2018Inflation,Kehagias:2013mya}.

With the aim of showing a scenario where emergence from a non-singular state is generic during expansion, we then considered a fluid model with interacting dark energy and dark matter. In this case, it is not necessary to have a quantum gravity era at high energy to avoid a singularity, e.g.\ see \cite{Lehners2019,Lehners2024}. Instead we have a classical de Sitter vacuum-dominated era at high energy from which our models emerge from, which is a past attractor for flat and open models, or a transition phase through a bounce for a subset of closed models. However, some models are cyclic and do not necessarily become close to this de Sitter phase. We also highlighted the models of interest that evolve with an early- and late-time acceleration connected by a decelerated period, which include open, flat and closed models. Therefore, at least qualitatively, these models are consistent with the observable Universe and emerge from a non-singular de Sitter state. 

For FLRW models that have a singularity, it is well known that this is very special in that it is matter dominated, while in more general GR models the shear anisotropy becomes dominant in approaching the singularity, which is said to be velocity dominated, e.g.\ see  \cite{Wainwright1997} and refs.\ therein. More in general, especially for bouncing models with a contracting phase, the question is if the singularity avoidance found in the FLRW case is stable against shear anisotropy \cite{Bozza2009}. In a long-wavelength approximation to inhomogeneities \cite{Landau1975}, this question can be  investigated by generalising non-singular FLRW models with a bounce to anisotropic Bianchi IX models, see \cite{Ganguly2019} and refs.\ therein. This will be the goal of our future investigation, generalising  the non-singular  scenario presented here based on dark matter interacting with dark energy, to study the stability of the results \cite{Burkmar2024}.

\newpage

\bibliographystyle{unsrt}
\bibliography{refs}

\end{document}